\documentclass[
aps,prd,
12pt,
nopreprintnumbers,
showpacs,
eqsecnum,
nofootinbib
]{revtex4-1}

\usepackage{graphicx}

\def\Tr{{\rm Tr\,}}
\def\tr{{\rm tr\,}}

\def\v0{{\bf 0}}

\begin{document}

\title{
Flux vacua in DBI type Einstein-Maxwell theory
}
\author{Takuya Maki}\email[]{maki@jwcpe.ac.jp}
\affiliation{
Japan Women's College of Physical Education, 
Setagaya,~Tokyo 157-8565,~Japan}
\author{Nahomi Kan}\email[]{kan@yamaguchi-jc.ac.jp}
\affiliation{
Yamaguchi Junior College,
Hofu-shi, Yamaguchi 747--1232, Japan}
\author{Koichiro Kobayashi}\email[]{m004wa@yamaguchi-u.ac.jp}
\author{Kiyoshi Shiraishi}\email[]{shiraish@yamaguchi-u.ac.jp}
\affiliation{
Yamaguchi University,
Yamaguchi-shi, Yamaguchi 753--8512, Japan}
\date{\today}

\begin{abstract}
We study compactification of extra dimensions in a theory of
Dirac-Born-Infeld (DBI) type gravity. 
We investigate the solution for Minkowski spacetime with an $S^2$ extra
space as well as that for de Sitter spacetime ($S^4$) with an $S^2$ extra
space. 
They are derived by the effective potential method in the presence of the
magnetic flux on the extra sphere.
We also consider the higher dimensional generalization of the solutions.
We find that, in a certain model, the
radius of the extra space has a minimum value independent of the
higher-dimensional Newton constant in weak-field limit.
\end{abstract}



\maketitle

\section{Introduction
\label{sec1}}

Recently the models including the higher derivative terms are widely
studied as a modified version of Einstein gravity.%
\footnote{For recent works on the higher-derivative gravity, see
\cite{Nelson} for example, and references therein.}
Moreover,
various works are reported about compactification with an extra space
in the higher derivative gravity
(for example, \cite{Wetterich,MH,CGTW,PT}). 
As one of such generalizations, which comes from other context,
the Dirac-Born-Infeld (DBI) type gravity has been considered by
Deser and Gibbons~\cite{DG} and studied by many authors~\cite{W}.
It is expected that the nonlinear nature of the model may remove the
possible singularity of spacetime. In our recent work, we considered a
model of Weyl invariant Dirac-Born-Infeld (DBI) type gravity~\cite{KMS}.%
\footnote{A three dimensional Weyl-invariant DBI gravity in studied by
Dengiz and Tekin \cite{ntekin}.}  
This model contains the Weyl gauge field.  It is natural
to   put the gauge field into DBI type gravity, since originally DBI
theory aimed at relaxing the singularity of the electric field. 

In the present paper, we first think about the theory with
the massless gauge field in six (or higher) dimensions (thus in the flat
spacetime, it seems to be the DBI electromagnetism), and consider
compactification of the extra dimensions. The Lagrangian density of our
model is the following:
\begin{equation}
{\cal L}=-\sqrt{-\det(f^2 g_{MN}-\alpha_1 R_{MN}-\alpha_2 R
g_{MN}+\beta F_{MN})}+(1-\lambda)f^D\sqrt{-g}\,,
\label{1}
\end{equation}
where $f$ is a mass scale and $\alpha_1$, $\alpha_2$, $\beta$ and
$\lambda$ are dimensionless parameters. 
The Ricci tensor $R_{MN}$ and the scalar curvature $R$ are the
quantities calculated from the metric $g_{MN}$, where $M, N$ run over
$0,1,\dots,D-1$. The antisymmetric tensor
$F_{MN}=\partial_MA_N-\partial_NA_M$ is the field strength of the Maxwell
theory. Note that our choice of overall sign is determined as the theory
describes the DBI electromagnetism in the flat spacetime.

A modification of the original DBI gravity of Deser and Gibbons
occurs as the presence of the scalar curvature term in the square-root.
It is plausible 
to include this term because the decomposition
of rank two tensor consists of a traceless symmetric part, a trace part
and an asymmetric part. The trace part of the curvature tensor should be
taken as being an independent term.

Now we consider the higher-dimensional extension, namely, $D$-dimensional
DBI type gravity. We assume $D$ dimensional spacetime.
Using the well-known expansion for a square-root of determinant of
sum of the identity matrix $1$ and a general matrix
$A$
\begin{equation}
\sqrt{\det{(1+A)}}
=1+\frac{1}{2}\tr A
+\frac{1}{8}((\tr A)^2-2\tr A^2)
+\cdots\,,
\end{equation}
we find
\begin{equation}
{\cal
L}\approx\sqrt{-g}\left(\frac{\alpha_1+D\alpha_2}{2}f^{D-2}R-
\frac{\beta^2}{4}f^{D-4}F^{MN}F_{MN}-f^{D}\lambda
\right)\,,
\label{l2}
\end{equation}
at the leading order in
the small coupling limit ($\alpha_1,\alpha_2,\beta\ll 1$),
and then the model
yields the Einstein-Maxwell theory.
Note that in this limit the higher-dimensional Newton constant is
proportional to $\alpha_1+D\alpha_2$. 
We should
pay attention to a magnetic flux in the
extra dimensions. 
Compactification with the flux in the
Einstein-Maxwell theory  was investigated by Randjbar-Daemi, Salam and
Strathdee about three decades ago~\cite{RSS1}.
Therefore existence of a similar solution for flux compactification
is expected.

Again we should remember that there also exist higher-curvature terms as
well as non-minimal coupled terms of the curvature and the Maxwell tensor
fields, which occurs in the next or further order of the expansion of the
Lagrangian (\ref{1}). Thus the compactification of space is partly
affected by the higher order of the curvature as in the model of
higher-derivative gravity~\cite{Wetterich,MH,CGTW,PT}.
Therefore we expect the new aspects of compactification in 
the flux compactification in DBI gravity.


In the next section, we give the assumption for the spacetime with
compactification and magnetic flux in the extra space.
In
Sec.~\ref{sec3}, compactification with four-dimensional Minkowski
spacetime is investigated by means of the effective potential. In
Sec.~\ref{sec4}, The case of $S^4$ (which is very similar to de Sitter
spacetime) compactification is studied. The higher-dimensional case is
examined in Sec.~\ref{sec5}. The last section is devoted to the summary
and prospects.

\section{Configuration of spacetime metric and flux
\label{sec2}}
In this section, we consider the following DBI type Lagrangian for six
dimensional spacetime expressed as
\begin{equation}
{\cal L}=-\sqrt{-\det(f^2 g_{MN}-\alpha_1 R_{MN}-\alpha_2 R
g_{MN}+\beta F_{MN})}+(1-\lambda)f^6\sqrt{-g}\,,
\end{equation}
where $M, N$ run over $0, 1, 2, 3, 5$, and $6$.
Note that one can rewrite the Lagrangian as in the form
\begin{eqnarray}
{\cal L}&=&-\sqrt{-\det {\cal M}_{MN}}+(1-\lambda)f^6\sqrt{-g}
\nonumber \\
&=&-\sqrt{-{g}}\sqrt{\det {\cal M}^{M}{}_{N}}+(1-\lambda)f^6\sqrt{-g}\,,
\end{eqnarray}
where
\begin{equation}
{\cal M}^{M}{}_{N}=f^2 \delta^{M}{}_{N}-\alpha_1 R^{M}{}_{N}-\alpha_2 R
\delta^{M}{}_{N}+\beta F^{M}{}_{N}\,.
\end{equation}

Now we assume that the spacetime is described by a direct product of
four-dimensional spacetime and an extra space, {\it i.e.}, the line
element is written by
\begin{equation}
ds^2=g_{\mu\nu}^{(4)}dx^{\mu}dx^{\nu}+g_{mn}^{(2)}dx^mdx^n\,,
\end{equation}
where
$\mu, \nu=0, 1, 2, 3$ while $m, n=5, 6$.
We will omit the index within the parentheses which
indicates the dimension, as long as confusion would not occur.

We suppose that the four-dimensional spacetime is a maximally symmetric
space. Then the Ricci tensor of the spacetime
is expressed as
\begin{equation}
R_{\mu\nu}=
\frac{1}{4}R^{(4)} g_{\mu\nu}\,,
\end{equation}
where $R^{(4)}$ is the scalar curvature of the four-dimensional spacetime.
For the Minkowski spacetime, $R^{(4)}=0$.

We adopt $S^2$ as the extra space.
Then we find
\begin{equation}
R_{mn}=\frac{1}{2}R^{(2)}g_{mn}=\frac{1}{b^2}g_{mn}\,,
\end{equation}
where 
$R^{(2)}$ and  $b$ is the scalar curvature and the radius of the
two-sphere, respectively.

Then we suppose that the constant `magnetic' flux penetrates the extra
sphere, just as in the model of Randjbar-Daemi, Salam and
Strathdee (RSS)~\cite{RSS1}. Namely we set
\begin{equation}
F_{mn}=B\sqrt{g^{(2)}}\varepsilon_{mn}\,,
\end{equation}
where $g^{(2)}=\det g_{mn}$.
The totally antisymmetric symbol $\varepsilon_{mn}$ takes the value $1$
for
$(m, n)=(5, 6)$.
The strength of flux is rewritten as $B=\tilde{B}/b^2$ where $\tilde{B}$
is a constant determined from a topological number.
Note that despite of the nonlinearity of equation of motion for
the Maxwell field, the topological configuration is a solution of the
equation of motion (even if $g_{mn}$ depends on $x^\mu$).

Before substituting above ans\"atze into the Lagrangian, we prepare the
matrices appearing in the root symbol. They are a four by four matrix
\begin{equation}
{\cal M}^{\mu}{}_{\nu}=\left[f^2-(\frac{\alpha_1}{4}+\alpha_2)R^{(4)}-
\alpha_2\frac{2}{b^2}\right]
\delta^{\mu}{}_{\nu}\,,
\end{equation}
and a two by two matrix
\begin{equation}
{\cal M}^{m}{}_{n}=\left[f^2-\alpha_2
R^{(4)}-(\alpha_1+2\alpha_2)\frac{1}{b^2}\right]
\delta^{m}{}_{n}+\beta\frac{\tilde{B}}{b^2}\sqrt{g^{(2)}}g^{m\ell}\varepsilon_{\ell
n}\,.
\end{equation}

We can easily calculate the determinant of each matrix as%
\footnote{More generally, one can calculate matrix determinant by the
known formula (gathered in Appendix A).}
\begin{equation}
\det
{\cal M}^{\mu}{}_{\nu}=\left[f^2-(\frac{\alpha_1}{4}+\alpha_2)R^{(4)}-
\alpha_2\frac{2}{b^2}\right]^4\,,
\end{equation}
and
\begin{equation}
\det
{\cal M}^{m}{}_{n}=\left[f^2-\alpha_2
R^{(4)}-(\alpha_1+2\alpha_2)\frac{1}{b^2}\right]^2+
\beta^2\frac{\tilde{B}^2}{b^4}\,.
\end{equation}

Since the following simple factorization holds,
\begin{equation}
\det {\cal M}^{M}{}_{N}=\det
{\cal M}^{\mu}{}_{\nu}\det
{\cal M}^{m}{}_{n}\,,
\end{equation}
we get the reduced Lagrangian as
{\footnotesize
\begin{eqnarray}
{\cal L}_0&=&-\sqrt{-g^{(4)}}(4\pi
b^2)\left(
\sqrt{\left[f^2-\left(\frac{\alpha_1}{4}+\alpha_2\right)R^{(4)}-
\alpha_2\frac{2}{b^2}\right]^4\left\{\left[f^2-\alpha_2
R^{(4)}-(\alpha_1+2\alpha_2)\frac{1}{b^2}\right]^2+
\beta^2\frac{\tilde{B}^2}{b^4}\right\}}\right.\nonumber \\
& &\left.-(1-\lambda)f^6\right)\,.
\end{eqnarray}
}

In this model, the effective Newton constant $G$ in four dimensional
spacetime can be read from the expansion of the Lagrangian in the small
curvature limit,
\begin{equation}
{\cal
L}=\sqrt{-g^{(4)}}\left[{\rm const.}+\frac{1}{16\pi
G}R^{(4)}+\cdots\right]\,.
\end{equation}
Thus we find  $(16\pi G)^{-1}$ for the constant 
radius $b_0$ as
\begin{eqnarray}
\frac{1}{16\pi
G}&=&\frac{4\pi
b_0^2\left(f^2-\frac{2\alpha_2}{b_0^2}\right)}{\sqrt{\left(f^2-
\frac{\alpha_1+2\alpha_2}{b_0^2}\right)^2+
\frac{\beta^2\tilde{B}^2}{b_0^4}}}\left\{\left(\frac{\alpha_1}{2}+
2\alpha_2
\right)\left[\left(f^2-
\frac{\alpha_1+2\alpha_2}{b_0^2}\right)^2+
\frac{\beta^2\tilde{B}^2}{b_0^4}\right]\right.\nonumber \\
& &\hspace{5cm}+\left.\alpha_2
\left(f^2-\frac{2\alpha_2}{b_0^2}\right)\left(f^2-
\frac{\alpha_1+2\alpha_2}{b_0^2}\right)\right\}\,.
\end{eqnarray}
Incidentally, this becomes simple for $\alpha_2=0$ and one finds
\begin{equation}
\frac{1}{16\pi
G}=2\pi b_0^2{f^2}{\alpha_1}
\sqrt{\left(f^2-
\frac{\alpha_1}{b_0^2}\right)^2+
\frac{\beta^2\tilde{B}^2}{b_0^4}}\,.
\end{equation}

\section{The solution with four dimensional Minkowski spacetime
\label{sec3}}
In this section, we seek the solution for the four-dimensional flat
spacetime. Then the whole spacetime is  
$M_{4}\times S^2$, the direct product of the Minkowski space and
two-sphere with a constant radius $b_0$.

According to Wetterich~\cite{Wetterich}, we use the method of the
effective potential for a static solution, instead of solving the equation
of motion derived from the Lagrangian directly. We now define a potential
\begin{equation}
V(y)=
y\left\{
\sqrt{\left(1-
\frac{2\alpha_2}{y}\right)^4\left[\left(1-\frac{\alpha_1+2\alpha_2}{y}\right)^2+
\beta^2\frac{\tilde{B}^2}{y^2}\right]}-(1-\lambda)\right\}\,,
\end{equation}
which is chosen as ${\cal L}_0\propto -V(f^2b^2)$.

Then the equation motion and the stability condition are equivalent
to~\cite{Wetterich}
\begin{equation}
\left.\frac{d V}{d y}\right|_{y=y_0}=V(y_0)=0\,,
\label{yy}
\end{equation}
and
\begin{equation}
\left.\frac{d^2 V}{d y^2}\right|_{y=y_0}>0\,,
\end{equation}
for the solution $y=y_0$. To make the equations (\ref{yy}) simultaneously
satisfied, we must tune the value of $\lambda$ to be a
specific value $\lambda_0$.

If $\beta\tilde{B}=0$, one obtains two solutions $y_0=2\alpha_2$ and
$y_0=\alpha_1+2\alpha_2$. For both cases, however, we cannot get the
four-dimensional Einstein gravity for weak-field limit, because 
${\cal L}_0\propto -\sqrt{R^{(4)}}$ in such a case.
Thus we consider $\beta\tilde{B}\ne 0$, and then further scaling of
the variable and parameters become convenient. This yields a scaled
potential,
\begin{equation}
\frac{1}{|\beta\tilde{B}|}V(y)\rightarrow
U(Y)=Y\left\{
\sqrt{\left(1-
\frac{2A_2}{Y}\right)^4\left[\left(1-\frac{A_1+2A_2}{Y}\right)^2+
\frac{1}{Y^2}\right]}-(1-\lambda)\right\}\,,
\end{equation}
where $Y=y/|\beta\tilde{B}|$, $A_1=\alpha_1/|\beta\tilde{B}|$ and
$A_2=\alpha_2/|\beta\tilde{B}|$.

Finally we find the solution for
\begin{equation}
\left.\frac{d U}{d Y}\right|_{Y=Y_0}=U(Y_0)=0\,,
\qquad {\rm and} \qquad
\left.\frac{d^2 U}{d Y^2}\right|_{Y=Y_0}>0\,,
\end{equation}
is given by
\begin{eqnarray}
Y_0&=&\frac{1}{2(A_1+6A_2)}\left\{1+A_1^2+14A_1A_2+24A_2^2\right.
\nonumber \\
& &+\left.\sqrt{
\left[1+A_1^2+2A_2(A_1-2\sqrt{6})\right]
\left[1+A_1^2+2A_2(A_1+2\sqrt{6})\right]}\right\}\,,
\label{s1}
\end{eqnarray}
and
\begin{equation}
1-\lambda_0=\sqrt{\left(1-
\frac{2A_2}{Y_0}\right)^4\left[\left(1-\frac{A_1+2A_2}{Y_0}\right)^2+
\frac{1}{Y_0^2}\right]}\,,
\end{equation}
provided that $Y_0$ is real and positive. We also require a positive
Newton constant $G$.
In conclusion, the value of $A_1$ must be larger than
$-\sqrt{\frac{3}{2}}$ and the allowed region for $A_2$ is restricted as
\begin{eqnarray}
-\frac{A_1}{6}<A_2\le\frac{1+A_1^2}{2(2\sqrt{6}-A_1)}
\qquad &{\rm for}&\quad 
-\sqrt{\frac{3}{2}}<
A_1<2\sqrt{6}\,,
\nonumber \\
A_2>-\frac{A_1}{6}
\qquad &{\rm for}&\quad A_1\ge 2\sqrt{6}\,.
\end{eqnarray}

For $A_2>0$, another solution $Y_0=2A_2$ exists.
The effective potential at  $Y_0=2A_2$
turns out to be the global minimum, but this solution cannot lead to 
Einstein gravity in four dimensions.
We will not analyze the possibility of quantum tunneling between the
vacua in the present paper.
Hereafter we investigate some cases with specific values for
the parameters.

Let us turn to the solution (\ref{s1}). For $A_1, A_2\ll 1$, we find
\begin{equation}
Y_0=\frac{1}{A_1+6A_2}\,,
\end{equation}
or $y_0=\frac{|\beta\tilde{B}|}{A_1+6A_2}$.
The solution coincides with that of the RSS model with the action
(\ref{l2}).

For $A_2=0$ (then $\alpha_2=0$), we find
\begin{equation}
Y_0=A_1+\frac{1}{A_1}\,,
\end{equation}
\begin{equation}
1-\lambda_0=\frac{1}{\sqrt{1+A_1^2}}\,,
\end{equation}
for $A_1>0$.
It is interesting to see the minimal value of the radius of $S^2$
is $\frac{\sqrt{|\beta\tilde{B}|}}{f}$, which is independent of the value
of $\alpha_1$. 

The inverse of the Newton constant is then given by
\begin{equation}
\frac{1}{16\pi
G}=2\pi {f^2}|\beta\tilde{B}|^{2}\sqrt{1+A_1^2}\,.
\end{equation}
The squared ratio of the compactification scale and the four-dimensional
Planck length is
\begin{equation}
\frac{b_0^2}{
G}=32\pi^2 |\beta\tilde{B}|^{3}\sqrt{\frac{(1+A_1^2)^3}{A_1^2}}
\ge 32\pi^2 |\beta\tilde{B}|^{3}\frac{3\sqrt{3}}{2}\,.
\end{equation}
Comparing with the result of the RSS model,
\begin{equation}
\frac{b_0^2}{G}=32\pi^2|\beta\tilde{B}|^{3}
\frac{1}{A_1}\,,
\end{equation}
when the model is described by the Lagrangian (\ref{l2}) with
$\alpha_2=0$,
we find that, in our DBI gravity model, the compactification scale cannot
be extremely smaller than the Planck length, provided that
$|\beta\tilde{B}|\sim 1$.


\section{the solution for the spacetime $S^4\times S^2$
\label{sec4}}
In this section we consider the four-dimensional spacetime is maximally
symmetric and has positive curvature, such that
\begin{equation}
R_{\mu\nu\rho\sigma}=
\frac{R^{(4)}}{12} (g_{\mu\rho}g_{\nu\sigma}-g_{\mu\sigma}g_{\nu\rho})
=
\frac{1}{a^2} (g_{\mu\rho}g_{\nu\sigma}-g_{\mu\sigma}g_{\nu\rho})\,.
\end{equation}
This can be realized if four-dimensional manifold is four-dimensional
sphere, whose radius is $a$.
It is known that  $S^4$ is equivalent to de Sitter spacetime
in the sense of its maximal symmetry.
For modeling our expanding universe, it is worth to study such a
situation. In this section, we choose
$\alpha_2=0$ in our model, for simplicity.

In this time, we use the effective potential
\begin{equation}
U(X,Y)=X^2 Y\left\{
\sqrt{\left(1-\frac{3A_1}{X}\right)^4\left[\left(1
-\frac{A_1}{Y}\right)^2+
\frac{1}{Y^2}\right]}-(1-\lambda)\right\}\,,
\end{equation}
where $X, Y$ represent for $f^2a^2/|\beta\tilde{B}|,
f^2b^2/|\beta\tilde{B}|$, respectively. Then the static compactification
is given by $X_0, Y_0$, the solution of
\begin{equation}
\left.\frac{\partial U}{\partial X}\right|_{X=X_0,~Y=Y_0}
=\left.\frac{\partial U}{\partial Y}\right|_{X=X_0,~Y=Y_0}=0\,.
\label{xy}
\end{equation}

For a sufficiently large positive value of $\lambda$, we can find a
solution of (\ref{xy}), though the solution corresponds to the saddle
point of the potential $U$.

Now we turn to searching solutions for a large $X$, or a large radius $a$.
For this purpose, we consider a linear deviation from the solution of the
Minkowski compactification obtained in the previous section.
We therefore set 
\begin{equation}
\lambda=\frac{1}{\sqrt{1+A_1^2}}+{\it\Delta}\lambda\,,\qquad
Y=A_1+\frac{1}{A_1}+{\it\Delta} Y\,,
\end{equation}
and here it is assumed that $0<{\it\Delta}\lambda\ll 1/\sqrt{1+A_1^2}$ and
${\it\Delta} Y\ll 1/A_1$.
Then the solution is approximately given by
\begin{equation}
X_0\approx \frac{3 A_1}{\sqrt{1+A_1^2}}({\it\Delta}\lambda)^{-1}\,,\qquad
{\it\Delta} Y_0\approx
\left(\frac{1+A_1^2}{A_1^2}\right)^{3/2}{\it\Delta}\lambda\,.
\end{equation}

This result suggests that the de Sitter universe with an extra space can
be obtained in our model of the DBI gravity.

Finally, we need to comment on a singular solution when $\lambda=1$.
The solution is given by $X_0=3 A_1$ and an arbitrary value for $Y$.
The similar situation can appear
in very early universe, if some modification such as varying $\lambda$
is incorporated into the present model. 

\section{a higher dimensional generalization
\label{sec5}}

In this section, we consider $(4+2p)$-dimensional spacetime ($p=1, 2,
\dots$). The compactified spacetime is supposed to be
$M_4\times (S^2)^p$.
We consider that the extra space consists of $p$ copies of the
original extra space. 
The radius of the $i$-th sphere (where $i=1,\dots,p$) is denoted by $b_i$,
while the flux on each sphere is denoted by $\tilde{B}_i$.

We investigate the corresponding solution in the model with $\alpha_2=0$
for simplicity also in this section.

We should find the solution $Y_{10},\cdots , Y_{p0}$, satisfying
\begin{equation}
\left.\frac{\partial U}{\partial Y_1}\right|_{Y_1=Y_{10}}=
\left.\frac{\partial U}{\partial Y_2}\right|_{Y_2=Y_{20}}=\cdots =
\left.\frac{\partial U}{\partial Y_1}\right|_{Y_p=Y_{p0}}=0
\quad {\rm and} \quad U(Y_{10},Y_{20},\cdots Y_{p0})=0\,,
\end{equation}
where
\begin{equation}
U(Y_1,\cdots Y_p)=\left(\prod_{i=1}^p Y_i\right)\left\{
\sqrt{\prod_{i=1}^p\left[\left(1-\frac{\alpha_1}{|\beta\tilde{B}_i|Y_i}\right)^2+
\frac{1}{Y_i^2}\right]}-(1-\lambda)\right\}\,.
\end{equation}
Here $Y_i=f^2b_i^2/|\beta\tilde{B}_i|$.

The solution is shown as in the case with the single sphere,
and should be read as
\begin{equation}
Y_{i0}=\frac{\alpha_1}{|\beta\tilde{B}_i|}+
\frac{|\beta\tilde{B}_i|}{\alpha_1}\,,
\end{equation}
and $\lambda$ is chosen as
\begin{equation}
1-\lambda_0=\prod_{i=1}^p\frac{1}{\sqrt{1+
\frac{\alpha_1^2}{|\beta\tilde{B}_i|^2}}}\,.
\end{equation}

The stability condition, 
\begin{equation}
\left.\det\frac{\partial^2 U}{\partial
Y_i\partial Y_j}\right|_{Y_1=Y_{10},\cdots ,Y_p=Y_{p0}}>0\,,
\end{equation}
which is seemingly complicated, is automatically satisfied because
\begin{eqnarray}
& &\left.\frac{\partial^2 U}{\partial
Y_i^2}\right|_{Y_1=Y_{10},\cdots ,Y_p=Y_{p0}}>0\quad {\rm for~}
i=1,\dots,p\,,\\
& &\left.\frac{\partial^2 U}{\partial
Y_i\partial Y_j}\right|_{Y_1=Y_{10},\cdots
,Y_p=Y_{p0}}=0\quad{\rm for~} i\ne j,~ i,j=1,\dots,p\,.
\end{eqnarray}

In the RSS model originated from the Lagrangian (\ref{l2}) (with setting
$\alpha_2=0$, $D=4+2p$ and other settings), the problem of similar
compactification can be solved by utilizing the following effective
potential
\begin{equation}
U(Y_1,\dots,Y_p)=\left(\prod_{i=1}^p
Y_i\right)\left[\sum_{i=1}^p\left(-\frac{\alpha_1}{|\beta\tilde{B}_i
|Y_i}+
\frac{1}{2Y_i^2}\right)+\lambda
\right]\,,
\end{equation}
and we find
\begin{equation}
Y_{i0}=\frac{|\beta\tilde{B}_i|}{\alpha_1}\,,\qquad
\lambda_0=\sum_{i=1}^p\frac{\alpha_1^2}{2|\beta\tilde{B}_i|^2}\,.
\end{equation}
We cannot distinguish our model from the RSS model if
$\frac{\alpha_1}{|\beta\tilde{B}_i|}\ll 1$; the fact comes from, of
course, the possible expansion shown in Sec.~\ref{sec1} only for small
couplings.

\section{Summary and outlook
\label{summary}}
In the present paper, the compactification in the DBI gravity with flux in
the extra space has been investigated.
The parameter region which allows the compactification has been
revealed. We have shown that the small couplings attached to the
curvature realizes similar compactification to that of the RSS model.

An interesting dependence of the radius of the extra space on the
parameter $\alpha_1$ was found in our model with
$\alpha_2=0$. This will be of more importance
if we consider the parameter as a dynamical variable, or
we generalize our model to include the term such as
$\phi^2 R_{MN}$, where $\phi$ is a scalar degree of freedom.

The analysis on stability against perturbation of higher modes, which
deforms the spherical shape of the extra space, is important, even though
the analysis on those mode will be complicated because of the
higher-derivative terms in our model. This issue is left for future works.

Spontaneous compactification to a football-shaped internal space in the
presence of a brane~\cite{football} is also worth studying in the
framework of the DBI type gravity models, because the higher curvature
terms affect the geometrical aspects of conical or nearly conical points
on the compact space.

The cosmological evolution of scale factors in our model is an important
subject.
Since the effective potential has a finite value at $b=0$ in our DBI type 
model with $\alpha_2=0$, the initial state of the universe may located at
$b=0$.  The simple condition is suitable
for quantum cosmology,%
\footnote{The quantum cosmology of the RSS model was studied by
Halliwell~\cite{Halli}.}  although the derivative terms in our model make
the canonical approach very complicated.

\begin{acknowledgments}
This study is supported in part by the Grant-in-Aid of Nikaido Research 
Fund.
\end{acknowledgments}


\begin{appendix}


\section{Determinant in $D$ dimensions}

Suppose that a general $D\times D$ matrix $M$ is given.
The determinant of $M$ is expressed as a polynomial of
$\Tr M^p ~(p\le D)$.

For $D=2$, we find
\begin{equation}
\det M=\frac{1}{2}\left[(\Tr M)^2-
(\Tr M^2)\right]\,.
\end{equation}
For $D=3$, we find
\begin{equation}
\det M=\frac{1}{6}\left[(\Tr M)^3-3
(\Tr M)(\Tr M^2)+2(\Tr M^3)\right]\,.
\end{equation}
For $D=4$, we find
\begin{eqnarray}
\det M&=&\frac{1}{24}\left[(\Tr M)^4-6(\Tr M)^2(\Tr M^2)+3(\Tr
M^2)^2\right.\nonumber\\
& &\qquad\left. +
8(\Tr M)(\Tr M^3)-6(\Tr M^4)\right]\,.
\end{eqnarray}
For $D=5$, we find
\begin{eqnarray}
\det M&=&\frac{1}{120}
\left[(\Tr M)^5-10(\Tr M)^3(\Tr M^2)
+15(\Tr M)(\Tr M^2)^2\right.\nonumber \\
& &\qquad+20(\Tr M)^2(\Tr M^3)-20(\Tr M^2)(\Tr M^3)\nonumber \\
& &\qquad\left.-30(\Tr M)(\Tr M^4)+24(\Tr M^5)\right]\,.
\end{eqnarray}
For $D=6$, we find
\begin{eqnarray}
\det M&=&\frac{1}{720}
\left[(\Tr M)^6-15(\Tr M)^4(\Tr M^2)
+40(\Tr M)^3(\Tr M^3)\right.\nonumber \\
& &\qquad+45(\Tr M)^2(\Tr M^2)^2-90(\Tr M)^2(\Tr M^4)\nonumber \\
& &\qquad\left.-120(\Tr M)(\Tr M^2)(\Tr M^3)+144(\Tr M)(\Tr
M^5)\right.\nonumber \\
& &\qquad\left.-15(\Tr M^2)^3+90(\Tr M^2)(\Tr M^4)
+40(\Tr M^3)^2-120(\Tr M^6)\right]\,.
\end{eqnarray}

\end{appendix}


\bibliographystyle{apsrev4-1}

\end{document}